\documentclass[12pt]{iopart}

\usepackage{graphicx,bbm,amssymb,amsopn}

\newcommand{\un}[1]{{\underline{#1}}}
\newcommand{\bra}[1]{{\langle #1 \vert}}
\newcommand{\xbra}[1]{{\langle #1 \vert}}

\newcommand{\xket}[1]{{\vert #1 \rangle}}

\newcommand{\xbraket}[2]{\langle #1 \vert #2 \rangle}

\newcommand{\ave}[1]{{\langle #1\rangle}}

\newcommand{\ii}{ {\rm i} }
\newcommand{\dd}{ {\rm d} }

\newcommand{\RR}{\mathbb{R}}
\newcommand{\CC}{\mathbb{C}}
\newcommand{\cc}{ {\hat c} }
\newcommand{\bb}{ {\hat b} }
\newcommand{\aaa}{ {\hat a} }
\newcommand{\y}{{\rm y}}
\newcommand{\x}{{\rm x}}
\newcommand{\z}{{\rm z}}
\newcommand{\LL}{{\hat {\cal L}}}
\newcommand{\NN}{{\hat {\cal N}}}
\newcommand{\PP}{{\hat {\cal P}}}
\newcommand{\mm}[1]{{\mathbf{#1}}}
\def\GLp{\Gamma_+^{\rm L}}
\def\GRp{\Gamma_+^{\rm R}}
\def\GLm{\Gamma_-^{\rm L}}
\def\GRm{\Gamma_-^{\rm R}}
\def\tr{{\,{\rm tr}\,}}
\def\ad{{\,{\rm ad}\,}}
\def\one{\mathbbm{1}}
\def\re{{\,{\rm Re}\,}}
\def\im{{\,{\rm Im}\,}}

\def\ness{\xket{{\rm NESS}}}

\begin{document}

\title[A general method to solve master equations for quadratic open Fermi systems]
{Third quantization: a general method to solve master 
equations for quadratic open Fermi systems}

\author{Toma\v{z} Prosen}

\address{Department of physics, FMF, University of Ljubljana,
Jadranska 19, SI-1000 Ljubljana, Slovenia}

\date{\today}

\begin{abstract}
 The Lindblad master equation for an arbitrary {\em quadratic} system of $n$ fermions is solved explicitly
 in terms of diagonalization of a $4n \times 4n$ matrix, provided that all Lindblad bath operators are
 {\em linear} in the fermionic variables. The method is applied to the explicit construction of non-equilibrium steady states and the calculation of asymptotic relaxation rates in the far from equilibrium problem of heat and spin transport in a nearest neighbor Heisenberg $XY$ spin $1/2$ chain in a transverse magnetic field.
\end{abstract}


\maketitle

\section{Introduction}
 
Understanding time evolution of an open quantum system of many interacting particles is of primary importance in fundamental problems of quantum physics, such as decoherence \cite{zurek,zeh} 
and closely related quantum measurement problem \cite{neumann,schlos}, quantum computation 
\cite{chuang,casati}, or the problem of computation of {\em non-equilibrium steady states} (NESS) in quantum statistical mechanics \cite{araki,ruelle,jaksic}.
Even though application of the methods of Hamiltonian dynamical systems and ergodic theory to quantum systems out of equilibrium gives many promising results \cite{piere2,lee,prosenjpa}, the field of open quantum systems is still lacking non-trivial explicitly solvable models, as compared to studies of closed (isolated) quantum systems where we know a large body of the so-called {\em completely integrable} systems \cite{korepin,fadeev}.
Examples of explicitly solvable models of master equations for open quantum systems are limited to quite restricted models of a single particle, single spin or harmonic oscillators (see e.g.
\cite{petruccione,haake,alicki}).

In this paper we show that the generator of the master equation of a
general quadratic system of $n$ interacting fermions which are coupled to a general set of Markovian baths, specified in terms of Lindblad operators which are linear in the fermionic variables - the so called
quantum Liouville super-operator (or Liouvillean) - can be explicitly diagonalized in terms of $2n$ {\em normal master modes}, i.e. anticommuting super-operators which act on the Fock space of density operators.
This can be understood as a complex (non-canonical) version of the Bogoliubov transformation \cite{lieb} lifted on the operator space, and has very powerful consequences:
(i) The NESS of the master equation can be understood as the `ground state' normal mode of the Liouvillean, whereas the long time relaxation rate is given by the eigenmode closest to the real axis.
(ii) The covariance matrix of NESS can be computed explicitly in terms of the eigenvectors of
$4n\times 4n$ antisymmetric complex matrix. It can be used to completely express physical observables in NESS, such as particle/spin densities, currents, etc.

We demonstrate the power of this novel method by applying it to the problem of heat and spin transport far from equilibrium
in nearest neighbor Heisenberg XY spin $1/2$ chains subject to a transverse magnetic field.
As a result we reproduce {\em ballistic transport} in the {\em integrable} spatially homogeneous case (see e.g. \cite{zotos,saito,hartmann,mejia05,mejia07,dhar,prosenjpa} for related
recent studies of quantum thermal conductivity in one dimension),
and predict {\em ideally insulating} behaviour (at all temperatures) in a disordered case of spatially random interactions/field.
Apart from obtaining numerical results which go by far beyond what was so far accessible by direct numerical
solution of the many-particle Lindblad equation, either directly or by means of 
quantum trajectories \cite{petruccione},
we also obtain two notable analytical results in the spatially homogeneous (non-disordered) case: (i) We compute the spectral gap of the Liouvillean i.e. the rate of
of relaxation to the NESS and show that it scales with the inverse cube of the chain length. (ii) We construct
{\em evanescent} normal master modes of the Liouvillean, for long chains, by which we explain quantitatively the exponential falloff of energy density or temperature profiles near the bath contacts.

The paper is organized as follows. In section \ref{sect:method} we shall outline a general method for the
diagonalization of the Liouvillean super-operator for finite quadratic open Fermi systems and an explicit construction of NESS.
In section \ref{sect:trivia} we illustrate the method by working out a simple example of a single fermion or a two level
quantum system in a bath. In section \ref{sect:nontrivia} we demonstrate the
usefulness  of the new method by applying it to quantum transport in XY spin chains. In section \ref{sect:conc} we discuss possible alternative applications and generalizations
of the method and reach some conclusions.

\section{General method of solution for the Lindblad equation}

\label{sect:method}

The general master equation governing time evolution of the density matrix $\rho(t)$ of an open quantum system, preserving trace and positivity of $\rho$, can be written in the Lindblad form
 \cite{lindblad,alicki} as (we set $\hbar=1$)
\begin{equation}
\frac{\dd\rho }{\dd t} = \LL\rho :=
-\ii [H,\rho] + \sum_{\mu} \left(2 L_\mu \rho L_\mu^\dagger - \{L_\mu^\dagger L_\mu,\rho\} \right)
\label{eq:lind}
\end{equation}
where $H$ is a Hermitian operator (Hamiltonian), $[x,y]:=xy-yx$, $\{x,y\}:=xy+yx$, and $L_\mu$ are arbitrary operators representing couplings to different
baths (at possibly different values of thermodynamic potentials). We are now going to describe a general method of explicit solution of (\ref{eq:lind}) for a {\em quadratic} system of $n$ fermions (or spins $1/2$) with {\em linear} bath operators
\begin{eqnarray}
H &=& \sum_{j,k=1}^{2n} w_j H_{jk} w_k = \un{w} \cdot \mm{H}\, \un{w} \label{eq:hamil}  \\
L_\mu &=& \sum_{j=1}^{2n} l_{\mu,j} w_j = \un{l}_\mu \cdot \un{w}\  \label{eq:lindb}
\end{eqnarray}
where $w_j$, $j=1,2,\ldots,2n$, are abstract {\em Hermitian} Majorana operators satisfying the
anti-commutation relations
\begin{equation}
\{w_j,w_k\} = 2\delta_{j,k} \qquad j,k =1,2,\ldots, 2n
\end{equation}
and generate a Clifford algebra.
Thus, $2n \times 2n$ matrix $\mm{H}$ can be chosen to be antisymmetric
$\mm{H}^T = -\mm{H}$. Throughout this paper $\un{x}=(x_1,x_2,\ldots)^T$ will designate a vector (column)
of appropriate scalar valued or operator valued symbols $x_k$.

Two notable examples, to which our formalism is immediately applicable, are:
(i) canonical fermions $c_m$, $m=1,2,\ldots, n$,
\begin{equation}
w_{2m-1}= c_m + c_m^\dagger \qquad
w_{2m} = \ii(c_m- c^\dagger_m)
\label{eq:physfermi}
\end{equation}
or (ii) spins $1/2$ with canonical Pauli operators $\vec{\sigma}_m$, $m=1,2,\ldots, n$,
\begin{equation}
w_{2m-1} = \sigma^\x_m \prod_{m'<m} \sigma^\z_{m'} \qquad
w_{2m} = \sigma^\y_m \prod_{m'<m} \sigma^\z_{m'}
\label{eq:jordan}
\end{equation}

Here we are not concerned with physical criteria for the validity of the so-called Markovian approximation under which eq. (\ref{eq:lind}) is derived, so we shall make no assumptions on the smallness of the bath coupling constants
$l_{\mu,j}$. We merely consider the Lindblad equation (\ref{eq:lind}) as a possible parametrization of an important subset of {\em Markovian} completely positive quantum channels and demonstrate its complete solvability for quadratic systems. Note that generalization of our formalism
to {\em explicitly time dependent} Hamiltonians $H(t)$ and Lindblad operators $L_\mu(t)$,
generating more general and possibly non-Markovian open system dynamics, is straightforward.
See e.g. \cite{wolf} for a discussion of {\em Markovianity}. 

\subsection{Fock space of operators}

We begin by associating a Hilbert space structure $x \to \xket{x}$ to a linear $2^{2n}=4^n$ dimensional space ${\cal K}$ of operators, with a canonical basis $\xket{P_{\un{\alpha}}}$ with
\begin{equation}
P_{\alpha_1,\alpha_2,\ldots,\alpha_{2n}} := w_1^{\alpha_1}w_2^{\alpha_2}\cdots w_{2n}^{\alpha_{2n}} \qquad
\alpha_j\in\{ 0,1\}
\end{equation}
{\em orthonormal} with respect to an inner product
\begin{equation}
\xbraket{x}{y} = 2^{-n} \tr x^\dagger y
\end{equation}
The form of the canonical basis of the operator space ${\cal K}$
suggests that it is just a usual Fock space with an unusual physical interpretation.
Namely we can define the following set of $2n$ adjoint  {\em annihilation linear
maps} $\hat{c}_j$ over ${\cal K}$
\begin{equation}
\cc_j \xket{P_{\un{\alpha}}} = \delta_{\alpha_j,1} \xket{w_j P_{\un{\alpha}}}
\label{eq:defanih}
\end{equation}
and derive the actions of their Hermitian adjoints - the {\em creation linear maps} $\hat{c}^\dagger$,
$\xbra{P_{\un{\alpha}'}}\cc^\dagger_j\xket{P_{\un{\alpha}}} =
\xbra{P_{\un{\alpha}}}\cc_j\xket{P_{\un{\alpha}'}}^* =
\delta_{\alpha'_j,1} \xbraket{P_{\un{\alpha}}}{w_j P_{\un{\alpha}'}}^*=
\delta_{\alpha_j,0} \xbraket{P_{\un{\alpha}'}}{w_j P_{\un{\alpha}}} $:
\begin{equation}
\cc^\dagger_j \xket{P_{\un{\alpha}}} = \delta_{\alpha_j,0} \xket{w_j P_{\un{\alpha}}}
\end{equation}
Straightforward inspection  then shows that they satisfy the canonical anticommutation relations
\begin{equation}
\{\cc_j,\cc_k\} = 0 \qquad \{\cc_j,\cc_k^\dagger\} = \delta_{j,k} \qquad j,k=1,2,\ldots, 2n
\end{equation}
The key is now to realize that the quantum Liouville map $\LL$ defined by eqs. (\ref{eq:lind},\ref{eq:hamil},\ref{eq:lindb}) can be written as a quadratic form in {\em adjoint Fermi maps} $\cc_j,\cc^\dagger_j$ (or for short, {\em a-fermions}). \footnote{Throughout this paper Dirac's bra-ket notation shall be used only for
a Hilbert space ${\cal K}$ of physical operators, including density operators, in a sense of GNS construction although here {\em all} spaces will be {\em finite} dimensional.
Symbols with a {\em hat} shall designate {\em linear maps} over the operator space ${\cal K}$.
For instance, we note a key distinction between {\em physical fermions} $c_m$
(\ref{eq:physfermi}) and a-fermions $\hat{c}_j$ (\ref{eq:defanih}).
}
\\\\
First, we consider the unitary part of Liouvillean
\begin{equation}
\LL_0 \rho := -\ii [H,\rho]
\label{eq:exprunitary}
\end{equation}
Since ${\cal K}$ is a Lie algebra, one defines the {\em adjoint representation} of a Lie derivative for an arbitrary $A\in {\cal K}$ back on ${\cal K}$ as, $\ad A \xket{B} := \xket{[A,B]}$. It is now straightforward to compute the action of a Lie derivative of a product of two Majorana operators on an arbitrary basis element
\begin{eqnarray}
\ad w_j w_k \xket{P_\un{\alpha}} &=& \xket{w_j w_k P_\un{\alpha}} - \xket{P_\un{\alpha} w_j w_k} = \nonumber \\
&=& 2 (\delta_{\alpha_j,1}\delta_{\alpha_k,0} + \delta_{\alpha_j,0}\delta_{\alpha_k,1})\xket{w_j w_k P_\un{\alpha}} = \nonumber \\
&=& 2 (\cc_j \cc^\dagger_k + \cc^\dagger_j \cc_k) \xket{P_\un{\alpha}} =
2 (\cc^\dagger_j \cc_k - \cc^\dagger_k\cc_j) \xket{P_\un{\alpha}}
\end{eqnarray}
Extending this relation by linearity to an arbitrary element of ${\cal K}$, it follows that for an arbitrary quadratic Hamiltonian (\ref{eq:hamil}) its Lie derivative has a very similar quadratic form in a-Fermi maps
\begin{equation}
\LL_0 = -\ii \ad H = -4\ii\sum_{j,k=1}^{2n} \cc^\dagger_j H_{jk} \cc_k =
-4\ii\, \un{\cc}^\dagger\cdot \mm{H}\,\un{\cc}
\label{eq:unitary}
\end{equation}
It is worth stressing here that for an arbitrary (complex) matrix $\mm{H}$, $\LL_0$ (\ref{eq:unitary}) conserves the total number of  a-fermions $\NN := \sum_{j} \cc^\dagger_j \cc_j = \un{\cc}^\dagger\cdot\un{\cc}$, namely $[\LL_0,\NN] = 0$.
\\\\
Second, we consider the action of the Lindblad maps
\begin{equation}
\LL_\mu \rho := 2 L_\mu \rho L_\mu^\dagger - \{L_\mu^\dagger L_\mu,\rho\} =
\sum_{j,k=1}^{2n} l_{\mu,j} l_{\mu,k}^* \LL_{j,k}\rho
\label{eq:exprlindb}
\end{equation}
where we write
$\LL_{j,k}\rho := 2 w_j \rho w_k - w_k w_j \rho - \rho w_k w_j$.
Again we proceed by computing the actions of $\LL_{j,k}$ on elements of the canonical basis of operator Fock space ${\cal K}$.  In order to do so, it is crucial to observe that the question whether $w_j$ commutes
or anticommutes with $P_{\un{\alpha}}$ depends on the number of a-fermions
$|\un{\alpha}|:=\sum_{k=1}^{2n} \alpha_k$ in $\xket{P_{\un{\alpha}}}$, namely
$P_{\un{\alpha}}w_j = (-1)^{|\un{\alpha}|+\alpha_j} w_j P_{\un{\alpha}}$, and hence
\begin{equation}
\LL_{j,k} \xket{P_{\un{\alpha}}}
= \left[2(-1)^{|\un{\alpha}|+\alpha_k} w_j w_k - w_k w_j - (-1)^{\alpha_j+\alpha_k} w_k w_j\right]
\xket{P_{\un{\alpha}}}
\label{eq:Ljk1}
\end{equation}
Observing that
\begin{eqnarray}
\phantom{(-1)^{\alpha_j}}\xket{w_j P_{\un{\alpha}}} &=& (\cc^\dagger_j + \cc_j)\xket{P_{\un{\alpha}}} \\
(-1)^{\alpha_j} \xket{w_j P_{\un{\alpha}}} &=& (\cc^\dagger_j - \cc_j)\xket{P_{\un{\alpha}}} \\
(-1)^{|\un{\alpha}|} \xket{P_{\un{\alpha}}} &=& \exp(\ii\pi\NN) \xket{P_{\un{\alpha}}}
\end{eqnarray}
one derives from (\ref{eq:Ljk1}) the general expression for $\LL_{j,k}$
\begin{eqnarray}
\LL_{j,k} &=& \left(\hat{\one}+\exp(\ii\pi\NN)\right)\left(2 \cc^\dagger_j \cc^\dagger_k - \cc^\dagger_j \cc_k -
\cc^\dagger_k \cc_j\right) \nonumber \\
&+& \left(\hat{\one}-\exp(\ii\pi\NN)\right)\left(2 \cc_j \cc_k - \cc_j \cc^\dagger_k -
\cc_k \cc^\dagger_j\right)
\end{eqnarray}
Obviously, the maps $\LL_{j,k}$, and hence also the total Lindblad part of Liouvillean
$\sum_\mu\LL_\mu$, do not conserve the number of a-fermions. But they conserve its {\em parity}
i.e. the product of any two creation/annihilation a-Fermi maps commutes with the
parity operation $\PP = \exp(\ii\pi\NN)$, with respect to which the operator
space can be decomposed into a direct sum
${\cal K} = {\cal K}^+ \oplus {\cal K}^-$, and even/odd operator spaces are orthogonally
projected as ${\cal K}^{\pm} = (\hat{\one} \pm \exp(\ii\pi\NN)){\cal K}$. Thus the positive
parity subspace ${\cal K}^+$ is a linear space spanned by $\xket{P_{\un{\alpha}}}$ with {\em even}
$|\un{\alpha}|$. All the maps $\LL_{j,k}$ now act separately on ${\cal K}^\pm$, $\LL_{j,k} {\cal K}^\pm
\subseteq {\cal K}^\pm$. For example, the maps defined on even parity subspace are indeed quadratic
in a-fermions
\begin{equation}
\LL_{j,k}\vert_{{\cal K}^+} = 4 \cc^\dagger_j \cc^\dagger_k - 2 \cc^\dagger_j \cc_k -
2 \cc^\dagger_k \cc_j
\label{eq:lindbpart}
\end{equation}
In this paper we shall focus on physical observables which are
products of an {\em even} number of Majorana fermions -- operators $w_j$ -- so we shall in the following discuss only
Liouville dynamics on the subspace ${\cal K}^+$. The extension to the dynamics of {\em odd} observables should be straightforward.

Putting the results (\ref{eq:exprunitary},\ref{eq:unitary},\ref{eq:exprlindb},\ref{eq:lindbpart})
together we arrive at the final compact quadratic representation of the Liouvillean
$\LL_+ := \LL\vert_{{\cal K}^+}$
\begin{equation}
\LL_+ = -2\, \un{\cc}^\dagger\cdot(2\ii \mm{H} + \mm{M} + \mm{M}^T)\,\un{\cc}
+ 2\,\un{\cc}^\dagger\cdot (\mm{M}-\mm{M}^T)\,\un{\cc}^\dagger
\label{eq:liouv1}
\end{equation}
where $\mm{M}$ is a complex Hermitian matrix parametrizing the Lindblad operators
\begin{equation}
M_{jk} = \sum_{\mu} l_{\mu,j} l_{\mu,k}^*
\end{equation}

\subsection{Reduction to normal master modes}

Next we want to show that  the representation (\ref{eq:liouv1})
allows us to reduce it further by a linear transformation of the set of maps
$\{\cc_j,\cc_j^\dagger; j=1,2,\ldots,2n\}$
to {\em normal master modes} (NMM) in terms of which the complete spectrum of the Liouvillean, as well as its eigenvectors,  can be explicitly constructed; in particular the zero-mode eigenvector which is just the physically relevant NESS.

In fact we proceed in analogy to Lieb {\em et al.} \cite{lieb} and define $4n$ adjoint Hermitian Majorana
maps $\aaa_r = \aaa_r^\dagger$, $r=1,2,\ldots,4n$:
\begin{equation}
\aaa_{2j-1} = \frac{1}{\sqrt{2}}(\cc_j + \cc^\dagger_j) \qquad
\aaa_{2j} = \frac{\ii}{\sqrt{2}}(\cc_j - \cc^\dagger_j)
\label{eq:aexpr}
\end{equation}
satisfying the anti-commutation relations
\begin{equation}
\{\aaa_r,\aaa_s\} = \delta_{r,s}
\label{eq:clifa}
\end{equation}
in terms of which the Liouvillean (\ref{eq:liouv1}) can be rewritten
as
\begin{equation}
\LL_+ = \un{\aaa}\cdot\mm{A}\,\un{\aaa} - A_0 \hat{\one}
\label{eq:liouv2}
\end{equation}
where $\mm{A}$ is an antisymmetric complex $4n\times 4n$ matrix with entries
\begin{eqnarray}
A_{2j-1,2k-1} &=&-2\ii H_{jk}-M_{jk}+M_{kj} \nonumber \\
A_{2j-1,2k}   &=&  \phantom{wj}2\ii M_{kj} \nonumber \\
A_{2j,2k-1}   &=& -2\ii M_{jk} \nonumber \\
A_{2j,2k}     &=&-2\ii H_{jk}+M_{jk}-M_{kj}
\label{eq:explA}
\end{eqnarray}
$\hat{\one}$ is an identity map over ${\cal K}$ and $A_0$ is a scalar
\begin{equation}
A_0 = 2\sum_{j=1}^{2n} M_{jj} = 2\tr\mm{M}
\end{equation}
Obviously, the {\em bi-linear} Liouvillean (\ref{eq:liouv2}) cannot be brought 
to a normal
form with a linear {\em canonical} transformation since the matrix $\mm{A}$ -- which shall in the following be referred to as a {\em shape matrix} of Liouvillean -- is not
anti-Hermitian like in Hamiltonian systems. So we should proceed in more
general terms.

We first recall few facts about complex antisymmetric matrices of even dimension.
If $\un{v}$ is a {\em right} eigenvector $\mm{A}\un{v} = \beta \un{v}$ with complex eigenvalue $\beta$, then $\un{v}$ is also a {\em left} eigenvector with eigenvalue $-\beta$, $\mm{A}^T \un{v} = -\mm{A} \un{v} = -\beta \un{v}$.
Hence eigenvalues always come in pairs $\beta,-\beta$. Let as assume
that $\mm{A}$ can be {\em diagonalized}\,\footnote{It is not known at present 
whether explict form (\ref{eq:explA}) guarantees diagonalizability of any such $\mm{A}$. Note that one can construct certain types of complex antisymmetric 
matrices with degenerate eigenvalues which cannot be diagonalized \cite{semrl}.}, i.e. there exist $4n$ linearly
independent vectors $\un{v}_r,r=1,\ldots,4n$ with the corresponding eigenvalues
$\beta_1,-\beta_1,\beta_2,-\beta_2,\ldots,\beta_{2n},-\beta_{2n}$,
\begin{equation}
\mm{A}\un{v}_{2j-1} = \beta_j \un{v}_{2j-1} \qquad
\mm{A}\un{v}_{2j} = -\beta_j \un{v}_{2j}
\label{eq:eigv}
\end{equation}
ordered such that $\re\beta_1\ge \re\beta_2\ge \ldots \ge \re\beta_{2n} \ge 0.$
The $2n$ complex numbers $\beta_j$ shall be referred to as {\em rapidities}.
It is easy to check that we can always choose and normalize $\un{v}_r$
such that\footnote{For a non-degenerate rapidity spectrum $\{\beta_j\}$
the proof of this statement is a trivial consequence of
antisymmetry $\mm{A}=-\mm{A}^T$, whereas in case of degeneracies it can be shown that one can always choose appropriate linear combinations of eigenvectors.}
\begin{equation}
\un{v}_r \cdot \un{v}_s = J_{rs} \quad \textrm{where} \quad
\mm{J} := \sigma^{\x}\otimes \mm{I}_{2n} =
\pmatrix{0 & 1 & 0 & 0 & \cdots \cr
                 1 & 0 & 0 & 0 & \cdots \cr
                 0 & 0 & 0 & 1 & \cdots \cr
                 0 & 0 & 1 & 0 & \cdots \cr
\vdots &\vdots & \vdots & \vdots& \ddots \cr}
\label{eq:norm}
\end{equation}
Let $\mm{V}$ be $4n\times 4n$ matrix whose $r$th row is given by $\un{v}_r$,
$V_{rs} := v_{r,s}$. Then eqs. (\ref{eq:eigv},\ref{eq:norm}) rewrite as
\begin{eqnarray}
\mm{A}\mm{V}^T &=& \mm{V}^T\mm{D}\quad \textrm{where}\quad \mm{D} := {\rm diag}\{\beta_1,-\beta_1,
\beta_2,-\beta_2,\ldots,\beta_{2n},-\beta_{2n}\} \label{eq:AVVD} \\
\mm{V} \mm{V}^T &=& \mm{J}
\label{eq:VVJ}
\end{eqnarray}
Expressing $\mm{V}^T$ in terms of (\ref{eq:VVJ}) and plugging the result into
eq. (\ref{eq:AVVD}) we arrive at a very convenient
canonical form of a generic complex antisymmetric matrix $\mm{A}$
\begin{equation}
\mm{A} = \mm{V}^T \mm{\Lambda} \mm{V}\quad\textrm{where}\quad
\mm{\Lambda} = \mm{D}\mm{J} =
\pmatrix{0 & \beta_1 & 0 & 0 & \cdots \cr
                 -\beta_1 & 0 & 0 & 0 & \cdots \cr
                 0 & 0 & 0 & \beta_2 & \cdots \cr
                 0 & 0 & -\beta_2 & 0 & \cdots \cr
\vdots &\vdots & \vdots & \vdots& \ddots \cr}
\label{eq:canform}
\end{equation}

Now we apply decomposition (\ref{eq:canform}) to the Liouvillean (\ref{eq:liouv2})
\begin{equation}
\LL_+ = \un{\aaa}\cdot \mm{V}^T \mm{\Lambda}\mm{V}\un{\aaa}  - A_0\hat{\one} =
(\mm{V}\un{\aaa})\cdot\mm{\Lambda}(\mm{V}\un{\aaa}) - A_0\hat{\one}
\label{eq:liouv4}
\end{equation}
Let us define the NMM maps
$\un{\bb} := (\bb_1,\bb'_1,\bb_2,\bb'_2,\ldots,\bb_{2n},\bb'_{2n}) :=
\mm{V}\un{\aaa}$ or
\begin{equation}
\bb_j = \un{v}_{2j-1}\cdot\un{\aaa} \qquad
\bb'_j = \un{v}_{2j}\cdot\un{\aaa}
\label{eq:bexpr}
\end{equation}
We note that due to (\ref{eq:clifa},\ref{eq:norm}) NMM maps satisfy
{\em almost-canonical} anti-commutation relations
\begin{equation}
\{\bb_j,\bb_k\} = 0 \qquad \{\bb_j,\bb'_k\} = \delta_{j,k} \qquad \{\bb'_j,\bb'_k\}=0
\label{acar}
\end{equation}
namely $\bb_j$ could be interpreted as annihilation map and $\bb'_j$ as a
creation map of $j$th NMM, but we should note that $\bb'_j$ is in general {\em not} the
Hermitian adjoint of $\bb_j$ \cite{thomas}.
In terms of NMM the Liouvillean (\ref{eq:liouv4}) now achieves a very
convenient normal form
\begin{equation}
\LL_+ = -2\sum_{j=1}^{2n} \beta_j \bb'_j \bb_j - B_0 \hat{\one}
\label{eq:liouv3}
\end{equation}
where $B_0 = A_0 - \sum_{j=1}^{2n}\beta_j$. We shall later show that
the constant $B_0$ is in fact equal to $0$.

\subsection{Non-equilibrium steady states and a complete spectrum of the Liouvillean}

The Liouvillean can always be represented in terms of a large but finite $4^n \times 4^n$
matrix. We shall now outline the procedure of complete construction of its spectrum in terms of NMM which are easy to calculate in terms of diagonalization of $4n \times 4n$ matrix $\mm{A}$ as described in the previous subsection.

We proceed by constructing the Liouvillean `vacuum'.
From the representation (\ref{eq:liouv1}) it follows immediately
that $\xbra{1} = \xbra{P_{(0,0\ldots,0)}}$ is left-annihilated by $\LL_+$,
$\xbra{1}\LL_+ = 0$, or equivalently $\LL_+^\dagger \xket{1} = 0$.
So we have just shown that $0$ is always an eigenvalue of $\LL_+$,
hence there should also exist the corresponding right eigenvector
$\ness$, normalized as $\xbraket{1}{{\rm NESS}} = \tr \rho_{\rm NESS} = 1$,
which represents physical NESS, i.e. stationary solutions of the Lindblad equation
(\ref{eq:lind})
\begin{equation}
\LL_+\ness = 0
\end{equation}
Let us define NMM number maps as $\NN_j := \bb'_j \bb_j$. From eqs. (\ref{acar}) it follows that $\NN_j$ satisfy a projection property $\NN_j^2 = \NN_j$, so they are
diagonalizable since no nontrivial Jordan block could satisfy the projection 
property. Furthermore, $\NN_j$ are mutually commuting $[\NN_j,\NN_k]=0$, so 
they can be simultaneously diagonalized and there should exist
a vacuum state on which all $\NN_j$ have value 0.
It follows from the stability of completely positive evolution (\ref{eq:lind}) that
all eigenvalues $\lambda$ of $\LL_+$ should obey $\re\lambda \le 0$, and since by
assumption $\re\beta_j \ge 0$, $\xbra{1}$ and $\ness$ should be the left and right vacua
which are simultaneously annihilated by NMM maps
\begin{equation}
\xbra{1} \bb'_j = 0 \qquad \bb_j \ness = 0
\label{eq:anihb}
\end{equation}
and hence also $\NN_j\ness = 0$.
Thus we have also shown that the NMM representation (\ref{eq:liouv3}) is only consistent
if $B_0=0$ so we find an interesting sum rule for rapidities
\begin{equation}
\sum_{j=1}^{2n} \beta_j = 2\tr\mm{M}
\end{equation}

The complete excitation spectrum and the corresponding left/right eigenvectors of the
Liouvillean are given in terms of a sequence of $2n$ binary integers
(NMM occupation numbers)
$\un{\nu}=(\nu_1,\nu_2,\ldots,\nu_{2n})$, $\nu_j\in\{0,1\}$,
\begin{equation}
\xbra{\Theta^{\rm L}_\un{\nu}}\LL_+ = \lambda_{\un{\nu}} \xbra{\Theta^{\rm L}_\un{\nu}}
\qquad
\LL_+ \xket{\Theta^{\rm R}_\un{\nu}} = \lambda_{\un{\nu}} \xket{\Theta^{\rm R}_\un{\nu}}
\end{equation}
namely
\begin{eqnarray}
\lambda_\un{\nu} &=& -2\sum_{j=1}^{2n} \beta_j \nu_j \label{eq:eval}\\
\xbra{\Theta^{\rm L}_\un{\nu}} &=& \xbra{1}{\bb_{2n}}^{\nu_{2n}}\cdots {\bb_2}^{\nu_2}\,{\bb_1}^{\nu_1}
\qquad
\xket{\Theta^{\rm R}_\un{\nu}} = {\bb_1}^{'\nu_1}\,{\bb_2}^{'\nu_2} \cdots {\bb_{2n}}^{' \nu_{2n}} \ness \label{eq:evec}
\end{eqnarray}
where by construction, left and right eigenvectors satisfy the bi-orthonormality relation
$\xbraket{\Theta^{\rm L}_\un{\nu'}}{\Theta^{\rm R}_\un{\nu}} = \delta_{\un{\nu}',\un{\nu}}$.

\subsection{The main general results: uniqueness of NESS, rate of relaxation to NESS, and expectation values of physical observables}

Given a {\em physical observable} $X\in {\cal K}^+$ and an {\em arbitrary} initial state with a
density operator $\rho_0 \in {\cal K}$, the time dependent expectation value of $X$ can be written in terms of the spectral resolution of the Liouvillean,
\begin{equation}
\exp(t \LL_+) = \sum_{\un{\nu}} \exp(t \lambda_{\un{\nu}})
\xket{\Theta^{\rm R}_{\un{\nu}}}\bra{\Theta^{\rm L}_{\un{\nu}}}
\label{eq:propag}
\end{equation}
namely
\begin{equation}
\ave{X(t)} = \tr X\rho(t) = \tr\!\!\left[X\exp(t\LL_+)\rho_0\right] =
\sum_{\un{\nu}} \exp(t\lambda_{\un{\nu}})\xbra{\Theta^{\rm L}_{\un{\nu}}}\rho_0 X\xket{\Theta^{\rm R}_{\un{\nu}}}
\label{eq:timedeptexp}
\end{equation}
We remind the reader that $\LL_+$ correctly represents physical Liouvillean only
on the subspace ${\cal K}^+$ of operators with {\em even} number of a-fermions. However, since the dynamics is closed on ${\cal K}^+$ and test physical 
observable $X$ also belongs to ${\cal K}^+$ it follows that the component of
$\rho_0$ from ${\cal K}^-$ does not contribute to the expectation value 
$\ave{X(t)}$.

Given the exact and explicit constructions developed in this section we can now make the following rigorous and useful conclusions, assuming throughout that Liouvillean shape matrix $\mm{A}$ is diagonalizable:
\\\\
{\bf Theorem 1:} NESS of Lindblad equation (\ref{eq:lind}) is {\em unique} if and only if
the rapidity spectrum $\{\beta_j\}$ does not contain $0$, in our ordering convention, if $\beta_{2n} \ne 0$. In the opposite case, if we have $d \ge 1$
vanishing rapidities, then we have a $2^d$ dimensional convex set of non-equilibrium steady states which can be spanned with $\xket{\Theta^{\rm R}_{(0,\ldots,0,\nu_{1},\ldots,\nu_{d})}}$.
\\\\
{\bf Theorem 2:} An arbitrary initial state with a density operator $\rho_0 \in {\cal K}$ converges
with time to NESS if and only if all rapidities have {\em strictly positive} real parts, $\re \beta_j > 0$.
Then, the rate of exponential relaxation to NESS is given by the {\em spectral gap} $\Delta$ of the Liouvillean which
equals $\Delta=2\re\beta_{2n}$.
\\\\
{\bf Theorem 3:} Assume that the rapidity spectrum does not contain 0, i.e. $\beta_{2n}\ne 0$. Then the expectation value of any quadratic observable $w_j w_k$  in a (unique) NESS can be explicitly
computed as
\begin{eqnarray}
\ave{w_j w_k}_{\rm NESS} &=& \delta_{j,k} + \xbra{1}\cc_j \cc_k \ness = \label{eq:pairwise} \\
&=& \delta_{j,k} + \frac{1}{2}\sum_{m=1}^{2n} \biggl(
v_{2m,2j-1} v_{2m-1,2k-1} - v_{2m,2j} v_{2m-1,2k}  \nonumber \\
&& \qquad\qquad- \ii v_{2m,2j} v_{2m-1,2k-1} - \ii v_{2m,2j-1} v_{2m-1,2k}\biggr)
\label{eq:pairwise2}
\end{eqnarray}

The statements of theorems 1 and 2 simply follow from exact and explicit spectral decomposition
(\ref{eq:eval},\ref{eq:evec},\ref{eq:propag}).

The proof of theorem 3 is also straightforward:
The first expression (\ref{eq:pairwise}) follows from the definition of the annihilation maps (\ref{eq:defanih})
and the explicit representation of the density operator of NESS, $\rho_{\rm NESS}$, in the canonical basis
$P_{\un{\alpha}}$. The second, very useful equality (\ref{eq:pairwise2}) is then obtained
by expressing $\cc_j$ thru (\ref{eq:aexpr}) in terms of NMM maps (\ref{eq:bexpr})
and using the annihilation relations (\ref{eq:anihb}).

The quadratic correlator of theorem 3 covers many physically interesting observables such as densities or currents.
However if one needs expectation values of more general observables, e.g. an expectation value of a high order monomial
$\ave{P_{\un{\alpha}}}_{\rm NESS} =
\xbra{1}\cc^{\alpha_1}_{1}\cc^{\alpha_2}_2\cdots\cc^{\alpha_{2n}}_{2n}\ness
$, then one may use a {\em Wick theorem}
and rewrite it as a sum of products of  pair-wise contractions (\ref{eq:pairwise}).

\section{Trivial example: A single fermion in a bath}

\label{sect:trivia}

In order to illustrate the method and demonstrate convenience of the results derived in the previous section we first work
out a simple example of a single fermion $n=1$ (or equivalently, an arbitrary qubit, a two-level quantum system), in
a thermal bath. We take the most general single fermion Hamiltonian $H = -\ii h w_1 w_2 + {\rm const} =
2h c^\dagger c + {\rm const}'$ and the following bath operators (see e.g. \cite{haake,wich})
\begin{equation}
L_1 = \frac{1}{2}\sqrt{\Gamma_1} (w_1 - \ii w_2)  =
\sqrt{\Gamma_1} c \qquad
L_2 = \frac{1}{2}\sqrt{\Gamma_2} (w_1 + \ii w_2) =
\sqrt{\Gamma_2} c^\dagger
\label{eq:canlind}
\end{equation}
where the ratio of coupling constants determine the bath temperature $T$,
$\Gamma_2/\Gamma_1 = \exp(-2h/T)$.
Leaving out the details of a straightforward calculation, simply following the steps of the previous section, we arrive at the following shape matrix of the Liouvillean (\ref{eq:liouv2})
\begin{equation}
\mm{A} = -h\mm{R} + \mm{B}_{\Gamma_+,\Gamma_-} \qquad A_0 = \Gamma_+
\end{equation}
where
\begin{equation}
\mm{R} := \pmatrix{ 0 & 0 & 1 & 0 \cr
                  0 & 0 & 0 & 1 \cr
                 \!\!-1 & 0 & 0 & 0 \cr
                  0 &\!\!\!-1 & 0 & 0}
\quad
\mm{B}_{\Gamma_+,\Gamma_-} :=
\pmatrix{
0 & \frac{\ii}{2} \Gamma_+ &\!\!\!-\frac{\ii}{2}\Gamma_- & \frac{1}{2}\Gamma_- \cr
\!\!-\frac{\ii}{2}\Gamma_+ & 0 & \frac{1}{2}\Gamma_- & \frac{\ii}{2}\Gamma_- \cr
\frac{\ii}{2}\Gamma_- &\!\!\!-\frac{1}{2}\Gamma_- & 0 & \frac{\ii}{2}\Gamma_+ \cr
\!\!-\frac{1}{2}\Gamma_- &\!\!\!-\frac{\ii}{2}\Gamma_- &\!\!\!-\frac{\ii}{2}\Gamma_+ & 0}
\label{eq:canbath}
\end{equation}
and $\Gamma_\pm := \Gamma_2 \pm \Gamma_1$.
Further, we compute NMM rapidities $\beta_{1,2} = \frac{1}{2}\Gamma_+ \pm \ii h$ and the
eigenvector matrix
\begin{equation}
\mm{V} = \pmatrix{
\frac{\Gamma_-}{\Gamma_+} - 1 & \ii \frac{\Gamma_-}{\Gamma_+} + \ii &
 -\ii \frac{\Gamma_-}{\Gamma_+} + \ii &  \frac{\Gamma_-}{\Gamma_+} + 1 \cr
 -\frac{1}{4} & -\frac{\ii}{4} & -\frac{\ii}{4} & \frac{1}{4} \cr
 \frac{\Gamma_-}{\Gamma_+} + 1 &  \ii \frac{\Gamma_-}{\Gamma_+} - \ii &
  \ii \frac{\Gamma_-}{\Gamma_+} + \ii &  -\frac{\Gamma_-}{\Gamma_+} + 1 \cr
  \frac{1}{4} & \frac{\ii}{4} & -\frac{\ii}{4} & \frac{1}{4}}\qquad
\end{equation}
Then, using theorem 3 we compute the expectation value of occupation number
$\ave{c^\dagger c} = \frac{1}{2} - \frac{\ii}{2} \ave{w_1 w_2} = \Gamma_2/(\Gamma_1 + \Gamma_2)$
which is what we expect in canonical equilibrium.

\section{Non-trivial example: transport in quantum spin chains}

\label{sect:nontrivia}

Here we work out a physically more interesting example, namely we construct NESS for the
magnetic and heat transport of a Heisenberg XY spin $1/2$ chain, with arbitrary -- either
homogeneous or positionally dependent (e.g. disordered) -- nearest neighbour interaction
\begin{equation}
H =
\sum_{m=1}^{n-1} \left( J^\x_m \sigma^\x_m \sigma^\x_{m+1} + J^\y_m \sigma^\y_m \sigma^\y_{m+1}\right)
+ \sum_{m=1}^n h_m \sigma^\z_m
\label{eq:hamsc}
\end{equation}
which is coupled to {\em two} thermal/magnetic baths {\em at the ends} of the chain, generated by two pairs of
canonical Lindblad operators \cite{wich} (similar to (\ref{eq:canlind}))
\begin{eqnarray}
L_1 &=& \frac{1}{2}\sqrt{\Gamma_1^{\rm L}} \sigma^{-}_1 \qquad
L_3 = \frac{1}{2}\sqrt{\Gamma_1^{\rm R}} \sigma^{-}_n \nonumber \\
 L_2 &=& \frac{1}{2}\sqrt{\Gamma_2^{\rm L}} \sigma^{+}_1 \qquad
 L_4 = \frac{1}{2}\sqrt{\Gamma_2^{\rm R}} \sigma^{+}_n
\label{eq:bathsc}
\end{eqnarray}
where $\sigma^{\pm}_m = \sigma^\x_m \pm \ii \sigma^\y_m$ and $\Gamma^{{\rm L},{\rm R}}_{1,2}$ are
positive coupling constants related to bath temperatures/magnetizations, for example if spins were non-interacting the bath
temperatures $T_{{\rm L},{\rm R}}$ would be given with $\Gamma^{{\rm L},{\rm R}}_2/\Gamma^{{\rm L},{\rm R}}_1 =
\exp(-2h_{\rm 1,n}/T_{{\rm L},{\rm R}})$.

Applying the inverse of Jordan-Wigner transformation (\ref{eq:jordan}),
$\sigma^\x_m = (-\ii)^{m-1}\prod_{j=1}^{2m-1} w_j$,
$\sigma^\y_m = (-\ii)^{m-1}(\prod_{j=1}^{2m-2} w_j)w_{2m}$,
we rewrite (\ref{eq:hamsc},\ref{eq:bathsc}) in
terms of Majorana fermions
\begin{eqnarray}
H &=& -\ii \sum_{m=1}^{n-1}\left(J^\x_m w_{2m} w_{2m+1} - J^\y_m w_{2m-1}w_{2m+2}\right)
     -\ii \sum_{m=1}^n h_m w_{2m-1}w_{2m} \\
L_1 &=& \frac{1}{2}\sqrt{\Gamma_1^{\rm L}} (w_1 - \ii w_2)
\qquad
L_3 = -\frac{(-\ii)^n}{2}\sqrt{\Gamma_1^{\rm R}} (w_{2n-1} - \ii w_{2n})W \nonumber \\
 L_2 &=& \frac{1}{2}\sqrt{\Gamma_2^{\rm L}}  (w_1 + \ii w_2)
\qquad
L_4 = -\frac{(-\ii)^n}{2}\sqrt{\Gamma_2^{\rm R}} (w_{2n-1} + \ii w_{2n})W
\end{eqnarray}
where $W=w_{1}w_{2}\cdots w_{2n}$ is a Casimir operator which commutes with all the elements of the Clifford algebra
generated by $w_j$ and squares to unity $W^2=1$. Therefore, $W$ {\em does not affect} the action of bath
operators (\ref{eq:exprlindb}) where $L_\mu$ enter quadratically, so we find
\begin{eqnarray}
\LL_1 + \LL_2 &=& -\Gamma_+^{\rm L} (\cc^\dagger_1\cc_1 + \cc^\dagger_2\cc_2) -
2\ii \Gamma_-^{\rm L} \cc^\dagger_1\cc^\dagger_2 \nonumber \\
\LL_3 + \LL_4 &=& -\Gamma_+^{\rm R} (\cc^\dagger_{2n-1}\cc_{2n-1} + \cc^\dagger_{2n}\cc_{2n}) -
2\ii  \Gamma_-^{\rm R} \cc^\dagger_{2n-1}\cc^\dagger_{2n}
\end{eqnarray}
leading to the bath shape matrix (\ref{eq:canbath}) identical to the single fermion case (\ref{eq:canlind}).
Again, carefully following the steps of section \ref{sect:method}, we derive the Liouvillean in the form
(\ref{eq:liouv2}) with $4n\times 4n$ shape matrix, which we write in a {\em block tridiagonal} form in terms of
$4\times 4$ matrices as
\begin{equation}
\mm{A}=
\pmatrix{
\mm{B}_{\rm L} - h_1 \mm{R} & \mm{R}_1 & \mm{0} & \cdots & \mm{0} \cr
-\mm{R}^T_1 & - h_2 \mm{R} & \mm{R}_2 &  \ddots & \mm{0} \cr
\mm{0} & -\mm{R}^T_2 & -h_3 \mm{R} &  & \vdots \cr
\vdots & \ddots & & \ddots & \mm{R}_{n-1} \cr
\mm{0} & \mm{0} & \cdots & -\mm{R}^T_{n-1} & \mm{B}_{\rm R} - h_n \mm{R} }
\label{eq:bigA}
\end{equation}
and $A_0= \Gamma_+^{\rm L} + \Gamma_+^{\rm R}$, where
$\mm{B}_{\rm L}:=\mm{B}_{\Gamma_+^{\rm L},\Gamma_-^{\rm L}}$,
$\mm{B}_{\rm R}:=\mm{B}_{\Gamma_+^{\rm R},\Gamma_-^{\rm R}}$
(in terms of (\ref{eq:canbath})), with
$\Gamma_{\pm}^{{\rm L},{\rm R}} := \Gamma_2^{{\rm L},{\rm R}}\pm\Gamma_1^{{\rm L},{\rm R}}$,
and
\begin{equation}
\mm{R}_m := \pmatrix{ 0 & 0 & J^\y_m & 0 \cr
                    0 & 0 & 0 & J^\y_m \cr
                    -J_m^\x & 0 & 0 & 0 \cr
                    0 & -J_m^\x & 0 & 0 }
\end{equation}
We are not able to perform a complete diagonalization of the antisymmetric matrix $\mm{A}$ (\ref{eq:bigA}) of the general XY model analytically. For example, even in the spatially homogeneous case
$J^{\x,\y}_m \equiv J^{\x,\y}, h_m\equiv h$ it is not possible to proceed like in the classical harmonic oscillator chain where the corresponding matrix is a sum of a Toeplitz and a bordered matrix \cite{rieder}. Namely, in our case $\mm{A}$ is a sum of a {\em block Toeplitz} and
{\em block bordered} matrix and its explicit exact diagonalization remains an open problem.
However, we stress that even relying on numerical diagonalization of $\mm{A}$ yielding a set of
rapidities $\beta_j$ and properly normalized eigenvector matrix $\mm{V}$,  represents a dramatic progress with respect to previously existing numerical methods which needed diagonalization of matrices which were exponentially large in $n$.
We shall later derive some exact theoretical and analytical results, explaining results of exact numerical
computations, in the special case of a {\em homogeneous} transverse Ising chain (subsection \ref{sect:ti}),  and the case of a
{\em disordered} XY chain (subsection \ref{sect:disord}) for which we shall relate NMM to the problem of Anderson localization in one dimension,

Let us continue by discussing transport observables in the spin chain whose expectation values in NESS are easy to calculate. Note that the bulk Hamiltonian (\ref{eq:hamsc})
can be written in terms of the two-body {\em energy density} operator
\begin{equation}
\!\!\!\!\!\!\!\!\!\!\!\!\!\!\!\!\!\!\!\!\!\!\!\!\!\!\!\!\!\!
H_m =  -\ii J^\x_m w_{2m} w_{2m+1} + \ii J^\y_m w_{2m-1}w_{2m+2}
     -\frac{\ii h_m}{2}w_{2m-1}w_{2m}-\frac{\ii h_{m+1}}{2}w_{2m+1}w_{2m+2}
     \label{eq:hdens}
\end{equation}
as $H=\sum_m H_m$. One can derive the local {\em energy current} $Q_m=\ii[H_m,H_{m+1}]$ from
the {\em continuity equation}
\begin{equation}
(\dd/\dd t)\ave{H_m} = \tr H_m \dd\rho/\dd t = \ave{\ii[H,H_m]}
= -\ave{Q_{m}} + \ave{Q_{m-1}}
\label{eq:cont}
\end{equation}
where $Q_m := \ii[H_{m},H_{m+1}] $
\begin{eqnarray}
Q_m =
&&2\ii(2J^\y_m J^\x_{m+1} w_{2m-1}w_{2m+3} + 2J^\x_mJ^\y_{m+1} w_{2m}w_{2m+4} -
\nonumber\\
  &&-J^\y_m h_{m+1} w_{2m-1}w_{2m+1} - J^\x_m h_{m+1} w_{2m}w_{2m+2}- \nonumber\\
  &&-h_{m+1}J^\x_{m+1}w_{2m+1}w_{2m+3}-h_{m+1}J^\y_{m+1}w_{2m+2}w_{2m+4}) \label{eq:hcurr}
\end{eqnarray}
The validity of the above continuity equation (\ref{eq:cont}) depends on two assumptions only:
(i) All Lindblad operators $L_\mu$ {\em commute} with the density $H_m$ in the {\em bulk},
$2 \le m \le n-2$ (second equality sign), and
(ii) $[H_m,H_{m'}] = 0$ if $|m-m'| \ge 2$ (third equality sign).

Using eq. (\ref{eq:pairwise2}) of theorem 3 we can now compute NESS expectation
values of energy density $H_m$ and energy current $Q_m$, and also of somewhat simpler
{\em spin density}
\begin{equation}
\sigma^\z_m = -\ii w_{2m-1}w_{2m}
\label{eq:sdens}
\end{equation}
and {\em spin current}
\begin{equation}
S_m = \sigma^\x_{m}\sigma^\y_{m+1}-\sigma^\y_{m}\sigma^\x_{m+1} = -\ii w_{2m}w_{2m+2} -\ii
w_{2m-1}w_{2m+1}
\label{eq:scurr}
\end{equation}
which are all quadratic in $w_j$.
Note, however, that the spin density satisfies continuity equation
$(\dd/\dd t)\ave{\sigma^\z_m} = -\ave{S_{m}} + \ave{S_{m-1}}$ only in the isotropic case, when
$J^\x_m \equiv J^\y_m$.

\subsection{Homogeneous transverse Ising chain}
\label{sect:ti}

Here we limit our discussion to the spatially homogeneous case
$J_n^{\x,\y}\equiv J^{\x,\y},h_n\equiv h$. We shall show that in this case the eigenvalue problem
\begin{equation}
\mm{A}\un{v} =
\beta\un{v}
\label{eq:evA}
\end{equation}
for (\ref{eq:bigA}) can be most easily and elegantly treated if formulated in terms of an abstract
inelastic scattering problem in one dimension, with asymptotic solutions given in terms of free normal modes
for the infinite translationally invariant chain
$\un{v}=(\ldots, \un{u} \xi^{m-1},\un{u} \xi^{m}, \un{u} \xi^{m+1},\ldots)^T$, where $\xi$ is a complex {\em quasi--momentum} (Bloch) parameter and $\un{u}$ is a
4-dimensional amplitude vector satisfying the condition
\begin{equation}
(-\mm{R}^T_1 \xi^{-1} - h \mm{R} + \mm{R}_1 \xi - \beta \mm{I}_4) \un{u} = 0
\label{eq:freemode}
\end{equation}
and the baths playing the role of inelastic (absorbing) scatterers at the edges of a finite lattice.
The `elastic' (Hamiltonian) version of this trick has been used to compute temporal correlation functions in kicked Ising chain \cite{ProsenPTPS}.

The singularity condition for the free mode equation (\ref{eq:freemode}) results, for a general homogeneous $XY$ model, in eight {\em master bands}  -
different values of momenta $\xi$ for each value of the spectral parameter (rapidity) $\beta$. In order to simplify the discussion - which will still get rather involved -
we shall in the following restrict ourselves to the transverse Ising model $J^\x = J, J^\y=0$.  In this case we find just two master bands with simple dispersion relations
\begin{equation}
\xi_{\pm}(\beta) = \frac{h^2 + J^2 + \beta^2 \pm \omega(\beta)}{2 h J}
\quad
\omega(\beta):=\sqrt{(h^2+J^2+\beta^2)^2 - (2 h J)^2}
\label{eq:dispersion}
\end{equation}
but each band is doubly-degenerate, since the corresponding amplitude problem (\ref{eq:freemode}) has two linearly independent solutions
\begin{equation}
\!\!\!\!\!\!\!\!\!\!\!\!\!\!\!\!\!\!\!\!
\un{u}^\pm_{1}(\beta) = \pmatrix{
-h (h^2 - J^2+\beta^2 \pm \omega) \cr
0 \cr
\beta ( h^2 + J^2 + \beta^2 \pm \omega) \cr
 0
 } \quad
\un{u}^\pm_{2}(\beta) = \pmatrix{
0 \cr
-h (h^2 - J^2+\beta^2 \pm \omega)\cr
0 \cr
\beta ( h^2 + J^2 + \beta^2) \pm \omega)
\label{eq:freemodes}
 }
 \end{equation}
 Naively speaking, $\xi_-$ represents left moving and $\xi_+$ right moving free modes, each having two possible polarizations. Note that  $\xi_- \xi_+ = 1$.
 For a general complex $\beta$ we shall choose the branch of square root $\omega(\beta)$ (\ref{eq:dispersion})
 for which $|\xi_-| \le 1$.
 Let us now write the scattering problem on the  {\em left} bath in terms of an ansatz
 \begin{equation}
 \un{v} = \pmatrix{ \un{u}_{\rm L} \cr
 \psi_1^- \un{u}^-_1 + \psi_2^- \un{u}^-_2  + \psi_1^+ \un{u}^+_1 + \psi_2^+ \un{u}^+_2 \cr
 (\psi_1^- \un{u}^-_1 + \psi_2^- \un{u}^-_2)\xi_-  + (\psi_1^+ \un{u}^+_1 + \psi_2^+ \un{u}^+_2)\xi_+ \cr
 (\psi_1^- \un{u}^-_1 + \psi_2^- \un{u}^-_2)\xi_-^2  + (\psi_1^+ \un{u}^+_1 + \psi_2^+ \un{u}^+_2)\xi_+^2 \cr
 \vdots}
 \label{eq:scatL}
 \end{equation} where $\un{u}_{\rm L}$ represents
 a $4$-dimensional vector of left-most eigenvector components,
 $\psi^-_{1,2}$ are the amplitudes of  (known) incident free modes, and  $\psi^+_{1,2}$ are the amplitudes of the scattered, outgoing free modes.
 Plugging the scattering ansatz to the eigenproblem (\ref{eq:evA}), the first two rows of $\mm{A}$ (\ref{eq:bigA}) result in 6
 linearly independent equations for 6 unknowns $\psi^+_{1,2},\un{u}_{\rm L}$. Eliminating four variables $\un{u}_{\rm L}$ we finally arrive to the
 non-unitary $2 \times 2$ S-matrix
 \begin{equation}
 \pmatrix{
 \psi^+_1 \cr
 \psi^+_2} = \mm{S}^{\rm L} \pmatrix{
 \psi^-_1 \cr
 \psi^-_2} \label{eq:SL} \\
 \end{equation}
 with
 \begin{eqnarray}
 S^{\rm L}_{11} &=& \tau^{-1} \beta^2 (-(\GLp)^4+4 (\GLp)^2 (\beta^2-3h^2) -16 h (h J^2 + \ii \GLm \omega)) \nonumber \\
 S^{\rm L}_{12} &=& \tau^{-1} \beta ( (\GLp)^3 + 8\ii \GLm h \beta + 4 \GLp (h^2 - \beta^2)) (2\ii \omega) \nonumber \\
 S^{\rm L}_{21} &=& \tau^{-1} \beta ( (\GLp)^3 - 8\ii \GLm h \beta + 4 \GLp (h^2 - \beta^2)) (-2\ii\omega) \nonumber \\
 S^{\rm L}_{22} &=& \tau^{-1} \beta^2 (-(\GLp)^4+4 (\GLp)^2 (\beta^2-3h^2) -16 h (h J^2 - \ii \GLm \omega)) \label{eq:Sexplicit} \\
\tau &:=& (\GLp)^4 \beta^2 + 8 \beta^2 (h^4 + (J^2 + \beta^2)(J^2 + \beta^2 - \omega) + h^2 (2 \beta^2 - \omega)) \nonumber \\
&-& 2(\GLp)^2 (h^4 + J^4 + 3\beta^4 + J^2 (2 \beta^2-\omega) - \beta^2\omega + h^2 (\omega - 2 J^2 - 4\beta^2)) \nonumber
 \end{eqnarray}
Similarly, one can solve the scattering problem from the {\em right} bath with the scattering ansatz
\begin{equation}
\un{v} = \pmatrix{
 \vdots \cr
 (\psi_1^+ \un{u}^+_1 + \psi_2^+ \un{u}^+_2)\xi_+^{-2}  + (\psi_1^- \un{u}^-_1 + \psi_2^- \un{u}^-_2)\xi_-^{-2} \cr
 (\psi_1^+ \un{u}^+_1 + \psi_2^+ \un{u}^+_2)\xi_+^{-1}  + (\psi_1^- \un{u}^-_1 + \psi_2^- \un{u}^-_2)\xi_-^{-1} \cr
 \psi_1^+ \un{u}^+_1 + \psi_2^+ \un{u}^-_2  + \psi_1^- \un{u}^-_1 + \psi_2^- \un{u}^-_2 \cr
 \un{u}_{\rm R}}
\end{equation}
defining the right S-matrix
\begin{equation}
 \pmatrix{
 \psi^-_1 \cr
 \psi^-_2} = \mm{S}^{\rm R} \pmatrix{
 \psi^+_1 \cr
 \psi^+_2}
 \end{equation}
 Note that since the two directions of free modes (\ref{eq:freemodes}) do not have left-right symmetry an explicit expression for $S^{\rm R}$ is considerably
 more complicated than (\ref{eq:Sexplicit}) and shall not be written out here.
 We shall now show that there exist two qualitatively different types of NMM - complex rapidities $\beta$
solving (\ref{eq:evA}) for sufficiently {\em large} $n$.

\begin{figure}[!t]
\begin{center}
 \includegraphics[scale=1.2]{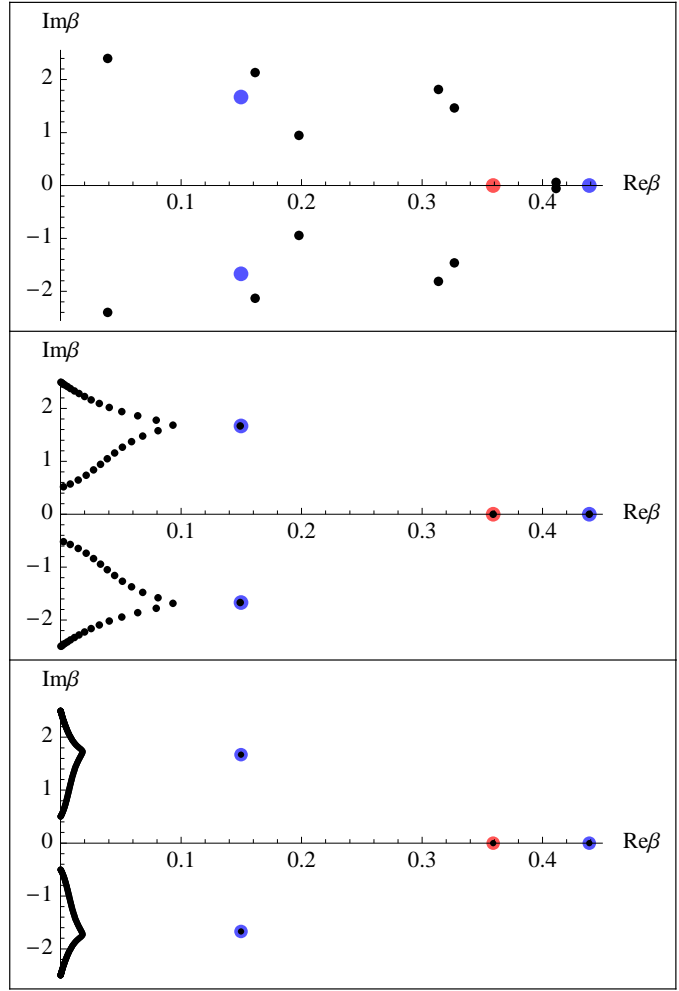}
\caption{
 Rapidities $\beta_j$ (black dots) for a transverse Ising chain with $J=1.5,h=1$ and bath couplings
 $\Gamma^{\rm L}_1=1,\Gamma^{\rm L}_2=0.6$,
 $\Gamma^{\rm R}_1=1,\Gamma^{\rm R}_2=0.3$, for three different sizes $n=6$ (upper), $n=30$ (middle), and $n=150$ (lower panel).
 Big blue/red dots indicate positions of evanescent rapidities (solutions of eq.(\ref{eq:4order}))  for the left/right bath.
\label{fig:rapid}}
\end{center}
\end{figure}

First, we shall discuss the so called {\em evanescent normal master modes}. These are characterized with amplitudes (\ref{eq:scatL}) which decay
exponentially with the distance from -- say the left -- bath, so the other -- the right boundary condition becomes physically irrelevant in the limit $n\to \infty$.
Such solutions $\psi^{+}_{1,2} = 0$ of eq. (\ref{eq:SL}) exist exactly when the determinant of S-matrix vanishes $\det S^{\rm L} = 0$. Using (\ref{eq:Sexplicit}) the
determinant can be written as $\det S^{\rm L} = (\beta/\tau)^2 p^{\rm L}(\beta)$ where\footnote{
Trivial zero $\beta=0$ of course does not represent a physical solution since then the whole S-matrix (\ref{eq:Sexplicit}) vanishes.}
\begin{eqnarray}
\!\!\!\!\!\!\!\!\!\!\!\!\!\!\!p^{\rm L}(\beta) &=& (\GLp)^8 \beta^2 - 4(\GLp)^6 ((h^2\!-\!J^2)^2  + (2J^2\!-\!4h^2)\beta^2 + 3\beta^4) \nonumber \\
&-& 16 (\GLp)^4(2 h^2 (h^2\!-\!J^2)^2 - (7h^4\!-\!6 h^2J^2\!+\!2 J^4)\beta^2 + 4(h^2\!-\!J^2)\beta^4 - 3 \beta^6)   \nonumber \\
&-& 64 (\GLp)^2(h^4 (h^2\!-\!J^2)^2-2h^2 J^4\beta^2 -(2h^4\!+\!4 h^2 J^2\!-\!J^4)\beta^4 + 2 J^2 \beta^6 + \beta^8) \nonumber \\
&+& 256 h^4 J^4 \beta^2 \label{eq:p}
\end{eqnarray}
Thus, for sufficiently large spin chains we find at most four NMM whose rapidities are given as the roots
of 4-th order polynomial in $\beta^2$
\begin{equation}
p^{\rm L}(\beta_{\rm evan}) = 0 \label{eq:evan}
\label{eq:4order}
\end{equation}
that are {\em not} simultaneously zeroes of $\tau(\beta)$.  Clearly, such NMM asymptotically do not depend on the chain size $n$.
In addition, we find evanescent NMM corresponding to the right bath simply by replacing $\GLp$ by $\GRp$ in (\ref{eq:evan},\ref{eq:p}).
In fig.~\ref{fig:rapid} we compare evanescent rapidities computed from eq. (\ref{eq:evan})  to numerically calculated spectrum
of $\mm{A}$,  at several different sizes $n$, for a typical case of transverse Ising chain,  $J=1.5,h=1.0$, strongly coupled to two baths at considerably different temperatures, $\Gamma^{\rm L}_1=1,\Gamma^{\rm L}_2=0.6$, $\Gamma^{\rm R}_1=1,\Gamma^{\rm R}_2=0.3$ Note that the same
parameter values will be used for numerical demonstrations throughout this subsection.

Second, we shall discuss the other extreme of {\em soft normal master modes} with rapidities that are closest to the imaginary axis, and thus determining the
spectral gap of the Liouvillean and relaxation time to NESS. Composing the scattering from the two baths with the free propagation along the chain (back and forth) we arrive at the general secular equation for the eigenvalue problem (\ref{eq:evA}) in terms of a $2\times 2$ determinant
\begin{equation}
\det ( \xi_+^{2(n-3)}  \mm{S}^{\rm R} \mm{S}^{\rm L}- \mm{I}_2) = 0
\label{eq:sec}
\end{equation}
In the absence of the baths, $\Gamma^{\rm L,R}_{\pm} = 0$, the solutions of the above problem exist only for real quasi-momenta,
namely $\xi_{\pm} = \exp(\pm\ii \vartheta), \vartheta \in \RR$. For such {\em extended} master modes the local coupling to the baths can be considered as a small
perturbation, thus only slightly perturbing the Bloch-like bands $\beta(e^{\ii\vartheta}) = \pm \ii \varepsilon(\vartheta)$ with `energy'
\begin{equation}
\varepsilon(\vartheta) = \sqrt{h^2 + J^2 - 2 |h J| \cos\vartheta}
\end{equation}
The softest NMM, namely the one for which the coupling to the baths is expected to be the weakest, should have nearly nodes at the ends of the chain, i.e.
$\vartheta \approx \pi/n$, or $\vartheta \approx \pi + \pi/n$,
and should thus lie near the band edges $\pm \ii |h| \pm \ii |J|$ (see fig.~\ref{fig:rapid}).
In the following we shall focus our calculation on the band edge
$\beta^* = \ii (|h| + |J|)$ which, as can be checked aposteriori by a straightforward but tedious calculation, always gives smaller
real part of the rapidity than the lower edge $\ii (|h|-|J|)$, and hence really determines the gap of the Liouvillean.
So we write
\begin{equation}
\beta = \ii(|h| + |J|) + z
\end{equation}
where $z\in\CC$ is a small parameter, and expand the S-matrices around the band edge
\begin{equation}
\mm{S}^{\rm L,R} = -\mm{I}_2 + \frac{4g}{\eta^{\rm L,R}}
\mm{Z}^{\rm L,R} \sqrt{-\ii z} + {\cal O}(|z|)
\label{eq:SZ}
\end{equation}
where $g:=\sqrt{\frac{|hJ|}{2(|h|+|J|)}}$, $\eta^{\rm L,R}:=(\Gamma^{\rm L,R}_+)^4 + 4 (\Gamma^{\rm L,R}_+)^2(4 h^2 + 2 |h J| + J^2) + 16 h^2 J^2$ and
\begin{eqnarray}
Z^{\rm L}_{11} &=& 4 |h|(\GLp)^2 + 16|h|(|h|+|J|)(|J|-\ii \GLm) \nonumber \\
Z^{\rm L}_{12} &=& -2 (\GLp)^3 - 16 \GLm |h|(|h|+|J|)-8(2h^2+2|hJ|+J^2) \nonumber\\
Z^{\rm L}_{21} &=& +2 (\GLp)^3 - 16 \GLm |h|(|h|+|J|)+8(2h^2+2|hJ|+J^2) \nonumber\\
Z^{\rm L}_{22} &=& 4 |h|(\GLp)^2 + 16|h|(|h|+|J|)(|J|+\ii \GLm) \label{eq:ZL}
\end{eqnarray}
and
\begin{eqnarray}
Z^{\rm R}_{11} &=& (\GRp)^4(2|h|+|J|)+4(\GRp)^2(8|h|^3+9h^2|J|+4|h|J^2+|J|^3) \nonumber \\
                          &+& 16 h^2|J|(|J|(3|h|+2|J|)-\ii\GRm(|h|+|J|)) \nonumber \\
Z^{\rm R}_{12} &=& -2 (\GRp)^3 - 16 \GRm |h|(|h|+|J|)-8(2h^2+2|hJ|+J^2) \nonumber\\
Z^{\rm R}_{21} &=& +2 (\GRp)^3 - 16 \GRm |h|(|h|+|J|)+8(2h^2+2|hJ|+J^2) \nonumber\\
Z^{\rm R}_{22} &=& (\GRp)^4(2|h|+|J|)+4(\GRp)^2(8|h|^3+9h^2|J|+4|h|J^2+|J|^3) \nonumber \\
                          &+& 16 h^2|J|(|J|(3|h|+2|J|)+\ii\GRm(|h|+|J|)) 
\label{eq:ZR}
\end{eqnarray}
Next we expand $\xi_+$ (\ref{eq:dispersion}) in $z$, yielding
\begin{equation}
\xi_+ = -1 - g^{-1} \sqrt{-\ii z} + {\cal O}(|z|)
\label{eq:xi1}
\end{equation}
and so the {\em free propagator} in (\ref{eq:sec}) can be written as
\begin{equation}
\xi_+^{2(n-3)} = \exp(2n g^{-1}\sqrt{-\ii z}) + {\cal O}(|z|).
\label{eq:xiexp}
\end{equation}
In eqs. (\ref{eq:SZ},\ref{eq:xi1},\ref{eq:xiexp})  the branch cut along the negative real axis has been chosen for $\sqrt{-\ii z}$.
Since the product of S-matrices in (\ref{eq:sec}) is near identity, the free propagator should be near one as well,
hence $2n g^{-1}\sqrt{-\ii z}$ should be near $2\pi\ii$. Let us define $z_0$ by setting $2n g^{-1}\sqrt{-\ii z_0} = 2\pi\ii$, so
\begin{equation}
z_0 = -\ii \pi^2 g^2 n^{-2}
\end{equation}
and write
$z = z_0 (1 + y) $ where $|y| \ll 1$ is another small complex parameter. However, since $z_0$ is purely imaginary, we need to
compute a small but non-vanishing $y$ which will, in the leading order in $n$, solve (\ref{eq:sec}) since
the real part of the soft mode's rapidity is determined as
\begin{equation}
\re \beta = \re z_0 y = \pi^2 g^2 n^{-2} \im y
\label{eq:y}
\end{equation}

\begin{figure}[!t]
\begin{center}

\vspace{2cm}

\hspace{3cm}  \includegraphics[scale=1]{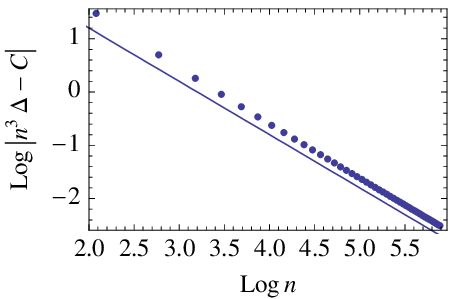}
\vspace{-5cm}

\includegraphics[scale=1.2]{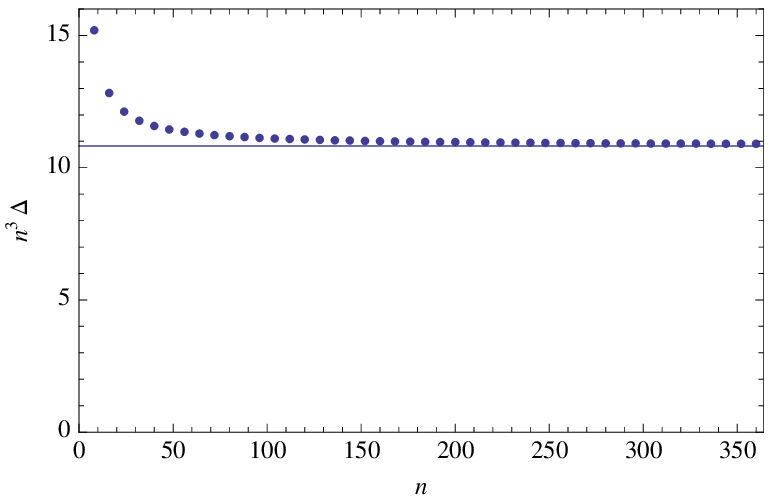}

\caption{
Spectral gap $\Delta$ times a third power of the chain length $n$ for a transverse Ising chain
with $J=1.5,h=1$ and bath couplings
 $\Gamma^{\rm L}_1=1,\Gamma^{\rm L}_2=0.6$,
 $\Gamma^{\rm R}_1=1,\Gamma^{\rm R}_2=0.3$.
Thin horizontal line indicates the theoretical
 asymptotic value (\ref{eq:delta}).
In the inset we show deviation from asymptotic constant value of $\Delta n^3$
in log-log scale and demonstrate that it decays as $\propto n^{-1}$ (thin line).
\label{fig:delta}}
\end{center}
\end{figure}

Now, writing $\sqrt{-\ii z} = \sqrt{-\ii z_0}\sqrt{1+y} = \ii \pi g n^{-1} (1 + y/2 - y^2/8) + {\cal O}(y^3)$ in (\ref{eq:SZ},\ref{eq:xiexp}),
plugging all that to eq. (\ref{eq:sec}) and computing to order ${\cal O}(n^{-2})$, noting that ${\cal O}(|z|)={\cal O}(n^{-2})$, we arrive to a
simple quadratic equation for $y$, whose solution, plugged to (\ref{eq:y}), gives the final result,
namely the sectral gap of Liouvillean $\Delta= 2\re \beta $
\begin{eqnarray}
\Delta &=& \frac{(2\pi h J)^2}{(|h|+|J|)^2} \frac{\Delta_1}{\Delta_2} n^{-3} + {\cal O}(n^{-4}) \label{eq:delta} \\
\Delta_1 &:=& 64 (\GLp + \GRp) h^2 J^2 (2 h^2 + 2|h J| + J^2)  \nonumber \\
                &+& 16 ((\GLp)^3 + (\GRp)^3) h^2 J^2  \nonumber \\
                &+& 16 \GLp \GRp (\GLp + \GRp) (2 h^2 + 2|hJ| + J^2)(4 h^2 + 2|hJ| + J^2) \nonumber \\
                &+& 4 \GLp \GRp ((\GLp)^3 + (\GLp)^2 \GRp + \GLp (\GRp)^2 + (\GRp)^3) (2 h^2+ 2|h J| + J^2) \nonumber \\
                 &+& (\GLp \GRp)^3 (\GLp+\GRp) \nonumber \\
\Delta_2 &:=&  ((\GLp)^4 + 4(\GLp)^2 (4 h^2 + 2|hJ| + J^2)+ 16 h^2 J^2 ) \nonumber \\
               &\times& ((\GRp)^4 + 4(\GRp)^2 (4 h^2 + 2|hJ| + J^2)+ 16 h^2 J^2) \nonumber
 \end{eqnarray}
 In fig.~(\ref{fig:delta}) we compare this analytical result to exact numerical calculations of
 the eigenvalue of $\mm{A}$ with minimal real part, confirming both, its precise numerical value and that
 the relative scaling of the next order correction is indeed ${\cal O}(n^{-1})$.

Note that, interestingly, both main analytical results of this subsection, namely evanescent and soft mode rapidities {\em do not depend} on
$\Gamma^{\rm L,R}_-$. Physically speaking, they only depend on the effective strengths of the bath couplings and not on the temperatures.

\begin{figure}[!t]
\begin{center}
 \includegraphics[scale=1.2]{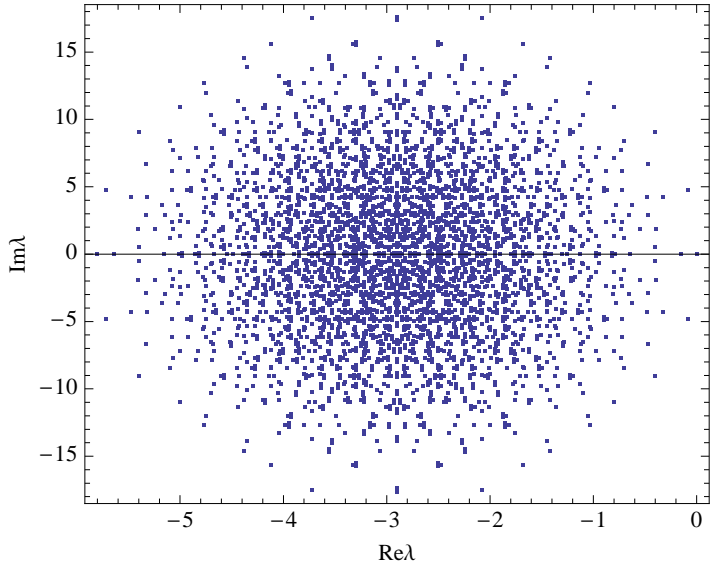}
\caption{
 Complete spectrum of $2^{12}$ complex eigenvalues of Liouvillean for a transverse Ising chain with $n=6$ spins and
 $J=1.5,h=1$ and bath couplings
 $\Gamma^{\rm L}_1=1,\Gamma^{\rm L}_2=0.6$,
 $\Gamma^{\rm R}_1=1,\Gamma^{\rm R}_2=0.3$ (the case of the upper panel of fig.~\ref{fig:rapid}).
\label{fig:spc}}
\end{center}
\end{figure}

\begin{figure}[!t]
\begin{center}
 \includegraphics[scale=1.2]{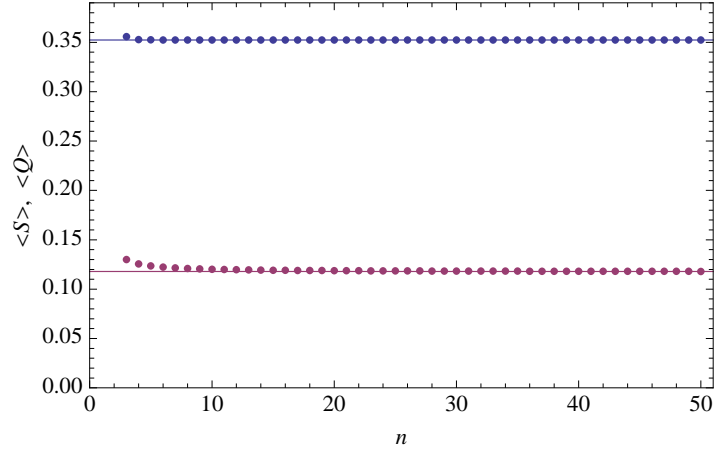}
\caption{
Energy current (upper/blue points), and average spin current (lower/red points), versus chain length
$n$ for a transverse Ising chain with  $J=1.5,h=1$ and bath couplings
 $\Gamma^{\rm L}_1=1,\Gamma^{\rm L}_2=0.6$,
 $\Gamma^{\rm R}_1=1,\Gamma^{\rm R}_2=0.3$.
\label{fig:current}}
\end{center}
\end{figure}

\begin{figure}[!t]
\begin{center}
 \includegraphics[scale=1.2]{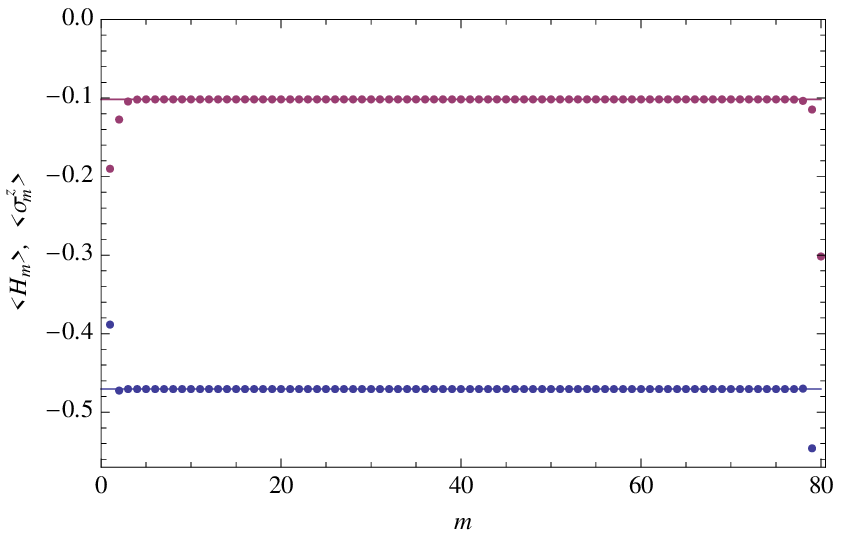}

 \vspace{-5cm}
 \hspace{10mm}\includegraphics[scale=1.05]{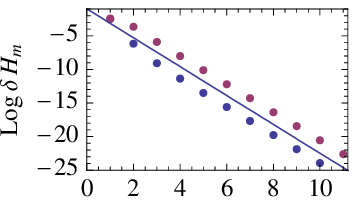}\hspace{-3mm}\includegraphics[scale=1.05]{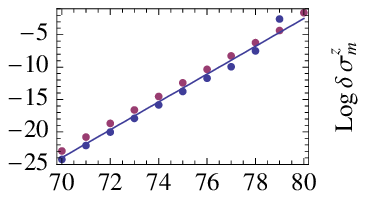}

 \vspace{2cm}
\caption{
Energy density profile (lower, blue points), and spin density profile (upper, red points), for a transverse Ising chain of $n=80$ spins
with  $J=1.5,h=1$ and bath couplings
 $\Gamma^{\rm L}_1=1,\Gamma^{\rm L}_2=0.6$,
 $\Gamma^{\rm R}_1=1,\Gamma^{\rm R}_2=0.3$. The insets display
logarithm of the difference to the bulk values
 $\delta H_m :=  |\ave{H_m} - H_{\rm bulk}|$ (blue points),
 $\delta \sigma^\z_m :=  |\ave{\sigma^\z_m} - \sigma^\z_{\rm bulk}|$
(red points) in comparison
 with $\pm (4 \log \xi_- ) m + {\rm const}$ with quasi-momentum $\xi_- = 0.584692$ corresponding 
(\ref{eq:dispersion})
 to the leading evanescent rapidity $\beta_{\rm evan} = 0.438739$ (full lines).
\label{fig:profile}}
\end{center}
\end{figure}

We end this  subsection by presenting some further numerical results on heat transport in the open transverse Ising chain in the Lindblad form.
In fig.~\ref{fig:spc} we demonstrate expression (\ref{eq:eval}) for constructing the full spectrum of the Liouvillean in terms of a set of
rapidities, for a short chain. In fig.~\ref{fig:current} we demonstrate eq. (\ref{eq:pairwise2}) of Theorem 3 by computing
the energy current $Q_m$ (\ref{eq:hcurr}), and the average spin current
$S=\frac{1}{n-1}\sum_{m=1}^{n-1} S_m$ (\ref{eq:scurr})
in NESS of a typical transverse Ising chain. Numerical results give a clear indication of {\em ballistic transport}
$\ave{Q} = {\cal O}(n^0), \ave{S}={\cal O}(n^0)$, however its rigorous proof and analytical calculation of the currents
would require full control over the complete set of NMM which is at present not available.
In fig.~\ref{fig:profile} we plot the energy density (\ref{eq:hdens}) and spin density (\ref{eq:sdens}) profiles in NESS.
Again, we note flat profiles in the bulk of the chain, $m,n-m \gg 1$, with exponential falloff due to adjustment to the non-equilibrium
bath values. Since the densities can be written, by means of (\ref{eq:pairwise2}), as $4-$point functions in NMM components,
the leading falloff exponents of the profile $|\ave{H_m} - H_{\rm bulk}| \sim
|\xi_-|^{4m} $ is given
by the quasi-momentum $\xi_-$ (\ref{eq:dispersion}) corresponding to the maximal evanescent rapidity $\beta_{\rm evan}$ (\ref{eq:evan}).

\subsection{Disordered XY chain}
\label{sect:disord}

In this subsection we treat the opposite extreme, a disordered XY chain (\ref{eq:hamsc}) where three sets of physical parameters are chosen
as {\em random uncorrelated} variables from {\em uniform} distributions on the intervals,
$J^\x_m \in [J^\x_{\rm min},J^\x_{\rm max}]$,
$J^\y_m \in [J^\y_{\rm min},J^\y_{\rm max}]$,
$h_m \in [h_{\rm min},h_{\rm max}]$. Clearly, the eigenvalue problem (\ref{eq:evA}) for the matrix (\ref{eq:bigA}) then becomes equivalent to the
Anderson tight-binding problem in one dimension for a quantum particle with a $4-$level internal degree of freedom.
We do not pursue any theoretical analysis of this problem here, but merely state that numerical investigations indicate existence of
exponential localization of {\em all} eigenvectors (or normal master modes) for disorder of any strength in anyone of system's parameters.

\begin{figure}[!t]
\begin{center}
 \includegraphics[scale=1.2]{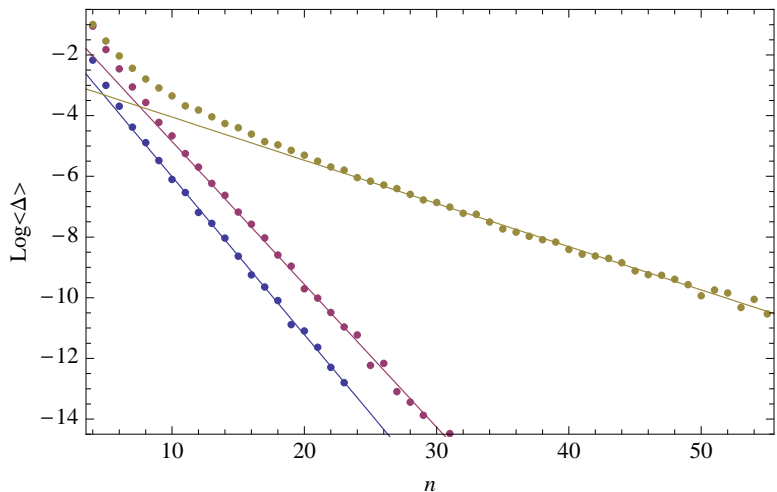}
\caption{
Average Liouvillean spectral gap $\ave{\Delta}$ versus the chain length
$n$ for disordered XY models:
(i) $J^\x_m = 0.5, J^\y_m = 0, h_m \in [1,2]$
(transverse Ising with field disorder, blue points),
(ii) $J^\x_m \in [0.5,2], J^\y_m = 0, h_m = 1$
(transverse Ising with interaction disorder, red points),
(iii) $J^\x_m \in [0.5,1],J^\y_m \in [0.5,1], h_m = 1$
(XY with interaction disorder, golden points),
all for bath couplings
 $\Gamma^{\rm L}_1=1,\Gamma^{\rm L}_2=0.6$,
 $\Gamma^{\rm R}_1=1,\Gamma^{\rm R}_2=0.3$.
 Full lines indicate exponential fits to right halves of data.
 Averaging is performed over $2000$ disorder realizations.
\label{fig:disdelta}}
\end{center}
\end{figure}

\begin{figure}[!t]
\begin{center}
 \includegraphics[scale=1.2]{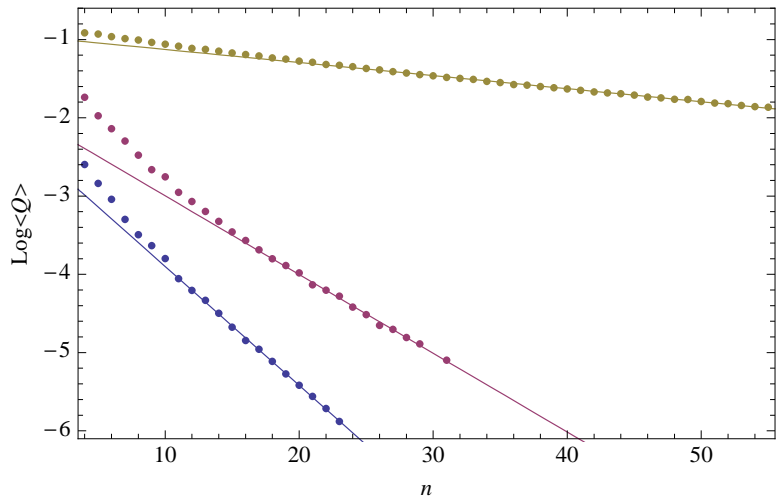}
\caption{
The scaling of the energy current $\ave{Q_m}$ with
chain length $n$ of the disordered XY model in the same regimes/parameters/plot styles as in fig.~\ref{fig:disdelta}.
\label{fig:discurrent}}
\end{center}
\end{figure}

\begin{figure}[!t]
\begin{center}
 \includegraphics[scale=1.2]{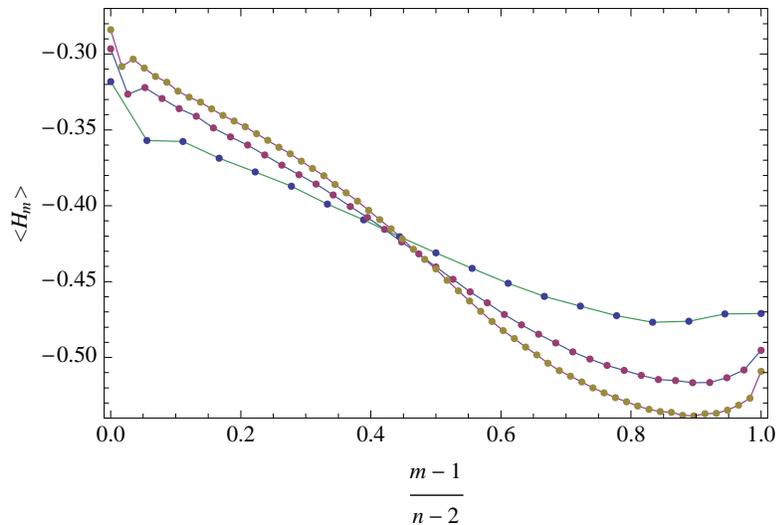}
\caption{
Scaled energy density profile of interaction disordered XY chains (case (iii) of fig.~\ref{fig:disdelta})
for three chain sizes: $n=20$ (blue points), $n=40$ (red points), $n=60$ (golden points).
Averages over 50000 disorder realizations have been performed.
\label{fig:disprofile}}
\end{center}
\end{figure}

With the picture of localization of  NMM in mind, the effect of the couplings to the heat baths at the chain's ends on quantum transport can be predicted by
theoretical arguments (see \cite{livi} for a review of related studies):
The spectral gap of the Liouvillean should be exponentially small $\Delta \sim \exp(-n/\ell)$ where $\ell$ is the localization length of NMM which is expected to be proportional to the square of inverse disorder strength.
This is demonstrated in fig.\ref{fig:disdelta}.
If all NMM are exponentially localized, the currents should decrease with the chain size $n$ faster than any power, perhaps exponentially, and the system should behave as an ideal insulator (at all temperatures). This is
demonstrated by straightforward numerical calculations of the heat current (\ref{eq:hcurr}) in fig.~\ref{fig:discurrent}.
In the final figure \ref{fig:disprofile}
we plot the energy density profile $\ave{H_m}$ (\ref{eq:hdens}) in a typical case of disordered XY chain,
versus a scaled spatial coordinate $(m-1)/(n-1) \in [0,1]$, for several different chain lengths $n$, and
demonstrate sharping up of energy density profiles with increasing $n$, which is again  indicating insulating behaviour.

\section{Discussion and conclusions}

\label{sect:conc}

The main result of the paper is a general method of explicit solution of master equations describing dynamics of open quantum system, under the condition that the system's Hamiltonian is {\em quadratic} and all Lindblad operators
are {\em linear} in canonical fermionic operators (which can either represent real physical fermions or any abstract
2-level quantum systems (qubits) thru the Jordan-Wigner transformation).
Using a novel concept of Fock space of physical operators (or density operators of physical states), and the adjoint structure of canonical creation and annihilation maps over this space, the problem can be treated in terms of a non-Hamiltonian generalization of the method of Lieb, Schultz and Mattis \cite{lieb} 
lifted to an operator space. We have explicitly constructed a non-canonical analog of Bogoliubov
transformation of the quantum Liouville map to normal master modes. 
Related ideas in the Hamiltonian context have been used by the author 
\cite{ProsenPTPS,prosen,pizorn} in order to approach the problem of real time dynamics and ergodic properties of {\em isolated} interacting many-body quantum systems.

As an illustration of the method we have solved far from equilibrium quantum heat and spin transport
problem in Heisenberg XY spin 1/2 chains which are coupled to canonical heat baths only at the two ends. Irrespectively of the strength of the coupling to the baths and their temperatures, we have shown a ballistic transport in the spatially homogeneous (non-disordered) case, and an ideally insulating behaviour in the disordered case associated to localization of normal master modes of the quantum Liouville operator. In this context the method can be considered as a simple alternative to the solution of quantum Langevin equations \cite{dhar}.

However, the method should easily be applicable to variety of other physical situations, for example if all fermions are coupled to the baths one could make a solvable model of genuine quantum diffusion, a many-body generalization
of the tight-binding model \cite{piere}. We also expect the method to be applicable
to the Redfield type of master equations (see e.g. \cite{piere}) - which do not conserve positivity for a short initial (slippage) time interval -
provided only the system part of the Hamiltonian is {\em quadratic} and system-bath couplings
are {\em linear} in fermionic variables.
Furthermore, extension of the method to open {\em many-boson} systems should be straightforward, simply by replacing anticommutators by commutators throughout the exposition of section \ref{sect:method}.

Treating density operators of NESS as elements of a Hilbert space of operators one may also extend the concept of {\em entanglement entropy}, with respect to a bipartition of a system of many fermions \cite{latorre}, to NESS  which can in our approach be viewed as a kind of ground state of the Liouvillean. Saturation of such {\em operator space entanglement entropy} \cite{pizorn} (which is suggested by numerical experiments \cite{PZ08}) indicates {\em efficient simulability} of NESS  by elaborate numerical methods such as {\em density matrix renormalization group} \cite{dmrg},
perhaps even for more general, non-solvable quantum systems.

As last we mention a more ambitious extension of the present work: Namely we propose to explore a question, whether more involved algebraic methods of solution of interacting many-body quantum systems, like e.g. Bethe Ansatz or quantum inverse scattering \cite{korepin}, could be generalized to open quantum systems, e.g. by means of the proposed concept of Fock space of operators. Could one discuss completely integrable open quantum systems which go beyond quadratic Liouvilleans?

\section*{Acknowledgements}

I gratefully acknowledge stimulating discussions with Pierre Gaspard, Keiji Saito and Walter Strunz, 
thank Carlos Mejia-Monasterio and Thomas H. Seligman for reading the manuscript and many
useful comments, and Iztok Pi\v zorn and Marko \v Znidari\v c for collaboration on related projects. The work has been supported by the grants P1-0044 and J1-7347 of Slovenian research agency (ARRS). Explicit analytical calculations reported in subsection (\ref{sect:ti}) were assisted by {\em Mathematica} software package.


\section*{References}

\begin{thebibliography}{10}

\bibitem{zurek} W. H. Zurek, Rev. Mod. Phys. {\bf 75}, 715 (2003).

\bibitem{zeh} E. Joos, H. D. Zeh, C. Kiefer, D. Giulini, J. Kupsch and I.-O. Stamatescu,
{\em Decoherence and the Appearance of a Classical World in Quantum Theory},
(Springer, 2003).

\bibitem{neumann}
J. von Neumann, {\em Mathematical Foundations of Quantum Mechanics}, Trans. Robert T. Geyer. (Princeton University Press, Princeton 1955).

\bibitem{schlos} M. Schlosshauer, Rev. Mod. Phys. {\bf 76}, 1267 (2004).

\bibitem{chuang} M.~A. Nielsen and I.~L. Chuang,
{\em Quantum Computation and Quantum Information}
(Cambridge University Press, Cambridge 2000).

\bibitem{casati} G. Benenti, G. Casati and G. Strini,
{\em Principles of Quantum Computation and Information.
Volume I: Basic Concepts} (World Scientific, Singapore 2004);
{\em Volume II: Basic Tools and Special Topics} (World Scientific, Singapore 2007).

\bibitem{araki}  H. Araki and E. Barouch, J. Stat. Phys. {\bf 31}, 327 (1983); \\
H. Araki, Publ. RIMS Kyoto Univ. {\bf 20}, 277 (1984).

\bibitem{ruelle} D. Ruelle, J. Stat. Phys. {\bf 98}, 57 (2000).

\bibitem{jaksic}
V. Jak\v si\v c and C.-A. Pillet, J. Stat. Phys. {\bf 108}, 787 (2002); Commun. Math. Phys. {\bf 226}, 131 (2002);\\
W. Aschbacher, V. Jak\v si\v c, Y. Pautrat and C.-A. Pillet,
{\em Inroduction to non-equilibrium quantum statistical mechanics},
in {\em Open Quantum Systems III. Recent Developments} Lecture Notes in Mathematics, {\bf 1882} (2006), 1-66.

\bibitem{piere2} P. Gaspard, Prog. Theor. Phys. Suppl. {\bf 165}, 33 (2006); Physica A {\bf 369}, 201 (2006).

\bibitem{lee} M. H. Lee, Acta Physica Polonica B {\bf 38}, 1837 (2007);
Phys. Rev. Lett. {\bf 87}, 250601 (2001); 
Phys. Rev. Lett. {\bf 49}, 1072 (1982).

\bibitem{prosenjpa} T. Prosen, J. Phys. A: Math. Theor. {\bf 40}, 7881 (2007).

\bibitem{korepin} V.~E.~Korepin, N.~M.~Bogoliubov, and A.~G.~Izergin,
 {\em Quantum Inverse Scattering and Correlation functions}
 (Cambridge University Press, Cambridge 1997).

\bibitem{fadeev} L. Fadeev, P. Van Moerbeke and F. Lambert (Eds.),
{\em 	
Bilinear Integrable Systems: from Classical to Quantum, Continuous to Discrete},
(NATO ARW Proceedings),
Springer Series: NATO Science Series II: Mathematics, Physics and Chemistry, Vol 201 (2006).

\bibitem{petruccione} H.-P. Breuer and F. Petruccione, {\em The Theory of Open Quantum Systems}
(Oxford University Press, Oxford 2002).

\bibitem{haake} F. Haake, {\em Quantum Signatures of Chaos}, 2nd edition
(Springer, 2001).

\bibitem{alicki} R. Alicki and K. Lendi, {\em Quantum dynamical semigroups and applications}
(Springer, 2007).

\bibitem{lieb} E. H. Lieb, T. D. Schultz and D. C. Mattis, Ann. Phys. (New York) {\bf 16}, 407 (1961).

\bibitem{zotos} X. Zotos, F. Naef and P. Prelov\v sek, Phys. Rev. B {\bf 55}, 11029 (1997).

\bibitem{saito} 
K. Saito, S. Takesue and S. Miyashita, Phys. Rev. E {\bf 61}, 2397 (2000);\\
K. Saito, Europhys. Lett. {\bf 61}, 34 (2003).

\bibitem{hartmann} M. Michel, M. Hartmann, J. Gemmer and G. Mahler,  Eur. Phys. J. B.  {\bf 34}, 325 (2003);\\
M. Michel, G. Mahler and J. Gemmer, Phys. Rev. Lett. {\bf 95}, 180602 (2005);\\
M. Michel, J. Gemmer and G. Mahler, Int. J. Mod. Phys. B {\bf 20}, 4855 (2006);\\
J. Gemmer, R. Steinigeweg and M. Michel, Phys. Rev. B {\bf 73}, 104302 (2006).

\bibitem{mejia05} C.~Mejia-Monasterio, T.~Prosen and G.~Casati, Europhys. Lett. {\bf 72}, 520 (2005);\\
G.~Casati and C.~Mejia-Monasterio, e-print {\tt arXiv:0710.3500v1 [cond-mat.stat-mech]}.

\bibitem{mejia07}  C.~Mejia-Monasterio and H.~Wichterich, 
e-print {\tt arXiv:0709.1412v1 [cond-mat.stat-mech]}.

\bibitem{dhar} A. Dhar and D. Roy, J. Stat. Phys. {\bf 125}, 805 (2006);
see also e-print {\tt arXiv:0711.4318v1 [cond-mat.stat-mech]}.

\bibitem{lindblad} G. Lindblad, Commun. Math. Phys. {\bf 48}, 119 (1976).

\bibitem{wolf} M. M. Wolf, J. Eisert, T. S. Cubitt and J. I. Cirac,
e-print {\tt arXiv:0711.3172v1 [quant-ph]}.

\bibitem{semrl} P. {\v Semrl}, {\em private communication}.

\bibitem{thomas}
We note a similarity to the formalism of second quantization
with non-orthogonal orbitals introduced in:
M. Moshinsky and T. H. Seligman, Ann. Phys. (New York) {\bf 66}, 311 (1971).

\bibitem{wich} H. Wichterich, M. J. Herich, H. P. Breuer, J. Gemmer and M. Michel, Phys. Rev. E, {\bf 76} 031115 (2007).

\bibitem{rieder} Z. Rieder, J. L. Lebowitz and E. Lieb, J. Math. Phys. {\bf 8}, 1073 (1967).

\bibitem{ProsenPTPS}
       T. Prosen, Prog. Theor. Phys. Suppl. {\bf 139}, 191 (2000).

\bibitem{livi} S. Lepri, R. Livi and A. Politi, Phys. Rep. {\bf 377}, 1 (2003).

\bibitem{prosen}
       T. Prosen, Phys. Rev. E \textbf{60}, 1658 (1999).

\bibitem{pizorn} T. Prosen and I. Pi\v zorn,  Phys. Rev. A {\bf 76}, 032316 (2007).

\bibitem{piere} M.~Esposito and P. Gaspard, Phys. Rev. B {\bf 71}, 214302 (2005); J. Stat. Phys. {\bf 121}, 463 (2005).

\bibitem{latorre}
   G.~Vidal, J.~I.~Latorre, E.~Rico, and A.~Kitaev, Phys. Rev. Lett. {\bf 90}, 227902 (2003);\\
       J.~I.~Latorre, E.~Rico, and G.~Vidal, Quant.~Inf.~Comp.{\bf 4}, 48 (2004).

\bibitem{PZ08} T. Prosen and M. \v Znidari\v c, to be submitted (2008).

\bibitem{dmrg}
       S.~R.~White, Phys. Rev. Lett. \textbf{69}, 2863 (1992);\\
       U. Schollw{\" ock} and S.~R.~White, in G.~G. Batrouni, and D.~Poilblanc (eds.): Effective models for low-dimensional strongly correlated systems, p.155, AIP, Melville, New York (2006);\\
       G.~Vidal, Phys.~Rev.~Lett. {\bf 91}, 147902 (2003); {\em ibid.} {\bf 93}, 040502 (2004).





\end{thebibliography}
\end{document}